\documentclass[twocolumn,aps,pre,superscriptaddress,floatfix,longbibliography]{revtex4-1}
\usepackage{amsmath,amsfonts,amssymb,amsthm,graphics,graphicx,epsfig,bbm}
\usepackage[colorlinks=true,citecolor=blue,linkcolor=blue,urlcolor=blue]{hyperref}

\usepackage{graphics, graphicx}
\usepackage{subfigure}
\usepackage{dcolumn}
\usepackage{bm}
\usepackage{xcolor}
\usepackage{mathtools}
\usepackage{setspace}
\usepackage{hyperref}
\usepackage{orcidlink}
\usepackage{wrapfig}
\usepackage{sidecap}

\begin{document}

\title{Contact temporal network during motility induced phase separation}

\author{Italo Salas \orcidlink{0009-0008-5970-6768}}
\email{italo.salas@ug.uchile.cl}
\affiliation{Departamento de F\'isica, Facultad de Ciencias, Universidad de Chile, Santiago Chile.}

\author{Francisca Guzm\'an-Lastra\orcidlink{0000-0002-1906-9222}}
\email{fguzman@uchile.cl}
\affiliation{Departamento de F\'isica, Facultad de Ciencias, Universidad de Chile, Santiago Chile.}

\author{Denisse Past\'en\orcidlink{0000-0002-4857-1051}} 
\email{denissepasten@uchile.cl}
\affiliation{Departamento de F\'isica, Facultad de Ciencias, Universidad de Chile, Santiago Chile.}

\author{Ariel Norambuena\orcidlink{0000-0001-9496-8765}}
\email{ariel.norambuena@usm.cl}
\affiliation{Departamento de F\'isica, Universidad T\'ecnica Federico Santa Mar\'ia, Casilla 110 V, Valpara\'iso, Chile.}%

\date{\today}

\begin{abstract}
Motility-induced phase separation (MIPS) is a paradigmatic non-equilibrium transition in active matter, determined by the Péclet number and packing fraction. We investigate the single-phase and phase-separated regimes of MIPS using a complex network approach, where networks are constructed from particle interactions over finite time windows. In the single-phase (gas-like) regime, the degree distributions $P(k)$ exhibit Gaussian behavior and resemble those of random graphs. Plotting the location and height of the $P(k)$ peak reveals a universal curve across different Péclet numbers at fixed packing fraction. In the phase-separated regime, we analyze the dense and dilute phases independently. The $P(k)$ distributions unveil distinct collective dynamics, including -like behavior in the dense phase and the emergence of active solid-like structures at longer times. Clustering coefficients and average path lengths in both phases stabilize rapidly, indicating that short simulations are sufficient to capture essential network features. Overall, our results show that network metrics expose both universal and phase-specific aspects of active matter dynamics. Notably, we identify distinct and previously unreported topological structures arising in the dense and dilute phases within the MIPS regime.
\end{abstract}

\maketitle

\section{Introduction}

Active Brownian particles (ABP) serve as a valuable framework for investigating the non-equilibrium dynamics and statistical properties of self-propelled agents, whether they are living organisms or artificial entities~\cite{romanczuk2012active,bechinger2016active,cates2024active,tan2022odd}. In contrast to passive Brownian particles studied in random diffusive systems, ABP offers a new dimension of complexity because of their inherent particle self-propulsion mechanisms. In particular, these active particles undergo a distinctive phase transition known as Motility-Induced Phase Separation (MIPS).

In this phenomenon, purely repulsive interactions and persistent self-propulsion spontaneously drive the system into coexisting dense and dilute phases \cite{fily2012athermal,redner2013structure,stenhammar2014phase,theers2018clustering,weitz2015self,paoluzzi2022motility,caporusso2020motility}. MIPS highlights how activity alone can trigger phase transitions without the need for attractive forces, emphasizing fundamental distinctions between equilibrium and active systems. Despite significant advances in theoretical, computational, and experimental investigations \cite{stenhammar2014phase,de2023sequential,de2021active,cates2024active,gompper20202020}, critical aspects of MIPS, at high densities, such as microscopic dynamics, structural complexities, and the temporal evolution of these phase-separated states, remain not yet understood.

The rich out-of-equilibrium ABP dynamics motivates a topological understanding of emergent phenomena using the language of complex networks. In particular, self-organization and phase transitions can be understood in terms of the temporal connections between particles. The contact particle-particle interactions within this system are still a significant topic of interest, where the density effects on self-diffusive dynamics, contact time, and effective velocity were recently studied from the kinetic theory lens \cite{soto2024kinetic,soto2025self}. Phase separation and clustering have also been analyzed in terms of particle shape~\cite{theers2018clustering,weitz2015self}, inertia~\cite{caprini2022role}, chirality~\cite{levis2019simultaneous,ma2022dynamical}, hydrodynamic interactions~\cite{zottl2023modeling,theers2018clustering,tan2022odd,petroff2015fast}, and quorum sensing~\cite{ridgway2023motility}. In recent years, agent-based models such as ABP have been used to model contagion dynamics on top of compartmental ecological models~\cite{norambuena2020understanding,de2023sequential,forgacs2022using}, and temporal networks have been used to track burstiness or clusterization in this type of  system~\cite{richardson2015beyond,gorochowski2017behaviour,zhong2023burstiness,bhaskar2021topological}. 

In recent years, studies conducted by complex networks have been applied to many natural systems such as social interactions~\cite{newman2001,Kertesz,gonzalez2008understanding}, seismology~\cite{abe2006,telesca2012,pasten_martin2021,pasten2018}, space plasmas~\cite{suyal2014visibility,mohammadi2021complex,saldivia2024}, or biological interactions~\cite{thiery,barabasi2011,scabini,riley2007large,stockmaier2021infectious}, among others. The formalism of complex networks can reveal nontrivial behavior in systems due to their geometrical and topological foundations~\cite{newman2003,albert2002statistical}. Through complex network analysis, it is possible to characterize and describe a physical system based on its topological properties, as well as to identify the type of network according to the structural organization of connections between nodes, for example, a random network\cite{erdos1960}, a scale-free network\cite{barabasi1999}, or a small-world network\cite{watts1998}. By representing systems as networks of interconnected nodes, this approach enables a quantitative and qualitative analysis of the underlying interaction patterns and their temporal evolution. However, applying complex network analysis to active matter, particularly ABP systems undergoing phase transitions such as MIPS, remains largely unexplored. In this work, we bridge this gap by introducing a novel approach that leverages complex network analysis to characterize the temporal complex networks formed by ABPs during motility-induced phase separation.

We systematically construct networks based on particle interactions over varying time windows to investigate how local particle contacts evolve into global structural properties. Our analysis reveals that the network structure resembles a random graph with scale-free behavior in the single-phase regime, capturing features analogous to an active gas. Conversely, we identify distinct and previously unreported network topologies for the dense and dilute phases in the MIPS regime, uncovering nontrivial correlations and structural complexity.

\section{Active Brownian Particles (ABPs) and motility-phase-separation(MIPS)}
\label{sec.abp}
We consider a system of $N$ self-propelled disk-like particles with diameter $\sigma$ moving in a rectangular region with area $A=L^2$, under periodic boundary conditions. The position vectors $\mathbf{r}_i(t)=[x_i(t),y_i(t)]$ and orientation vectors $\hat{\mathbf{u}}_i(t)=
[\cos\theta_i(t),\sin\theta_i(t)]$ evolves according to overdamped Langevin dynamics on a flat two-dimensional surface
\begin{eqnarray}
\dot{\mathbf{r}}_i&=&u_{0}\hat{\mathbf{u}}_i(t)+(1/\gamma_t)(\boldsymbol{\xi}_{i,T}(t)-\nabla_{\mathbf{r}_i}U),  \label{eq1}\\
    \dot{\hat{\mathbf{u}}}_i&=&(1/\gamma_r)\boldsymbol{\xi}_{i,R}(t)\times\hat{\mathbf{u}}_i, \label{eq2}
\end{eqnarray}
where $\gamma_t,\gamma_r$ are the translational and rotational Stokes friction coefficients in the $x-y$ plane, respectively, and $u_0$ is the constant self-propelled velocity. In Eqs.\eqref{eq1} and \eqref{eq2}, $\boldsymbol{\xi}_{T}(t)$ and $\boldsymbol{\xi}_{R}(t)$ represent translational and rotational Gaussian white noises with zero mean, i.e. $\langle\boldsymbol{\xi}_{T}
(t)\rangle=\langle\boldsymbol{\xi}_{R}(t)\rangle=\mathbf{0}$, and time correlations given by $\langle\boldsymbol{\xi}_{T}(t_1)\boldsymbol{\xi}_{T}(t_2)\rangle=D_t\delta(t_1-t_2)$ and $\langle
\boldsymbol{\xi}_{R}(t_1)\boldsymbol{\xi}_{R}(t_2)\rangle=D_r\delta(t_1-t_2)$, where $D_t$ and $D_r$ are defined as translation and rotation diffusion parameters, respectively. 
The interaction potential $U(\{\mathbf{r}_i\})$ in Eq.~\eqref{eq1} corresponds to the summation over steric interactions between particles at time $t$ and is given by

\begin{equation}
    U=\sum_{i\neq j}U_{\text{WCA}}(\mathbf{r}_i,\mathbf{r}_j), \label{eq.3}
\end{equation}

 where $\mathbf{r}_j$ is the position of the $j-$th particle, and $U_{\text{WCA}}$ is the Weeks-Chandler-Andersen (WCA) pair potential 
 
\begin{equation}\label{eq.4}
U_{\text{WCA}}(\mathbf{r}_i,\mathbf{r}_j) =  \left\{\begin{array}{cc}
     \displaystyle{4 \varepsilon\left[\left(\frac{\sigma}{r_{i j}}\right)^{12}-\left(\frac{\sigma}{r_{ij}}\right)^{6}\right]},    &   r_{i j} \leq 2^{1/6} \sigma\\
     0    & \mbox{otherwise} 
    \end{array} \right.
\end{equation}

Here, $r_{ij} = |\mathbf{r}_i - \mathbf{r}_j|$ is the interparticle distance, $\sigma$ is the particle diameter, and $C_r = 2^{1/6} \sigma$ sets the cutoff for the interaction potential. The parameter $\varepsilon$ controls the interaction strength. Two dimensionless parameters govern the system: the packing fraction $\phi = [N\pi(\sigma/2)^2]/L^2$, which measures particle area relative to system size, and the Péclet number $\text{Pe} = 3u_0 / (D_r \sigma)$, which quantifies the ratio of active propulsion to rotational diffusion. MIPS emerges within specific ranges of $\phi$ and $\text{Pe}$~\cite{redner2013structure,levis2017active}.

\begin{figure}[ht]
\centering 
\includegraphics[width=\columnwidth ]{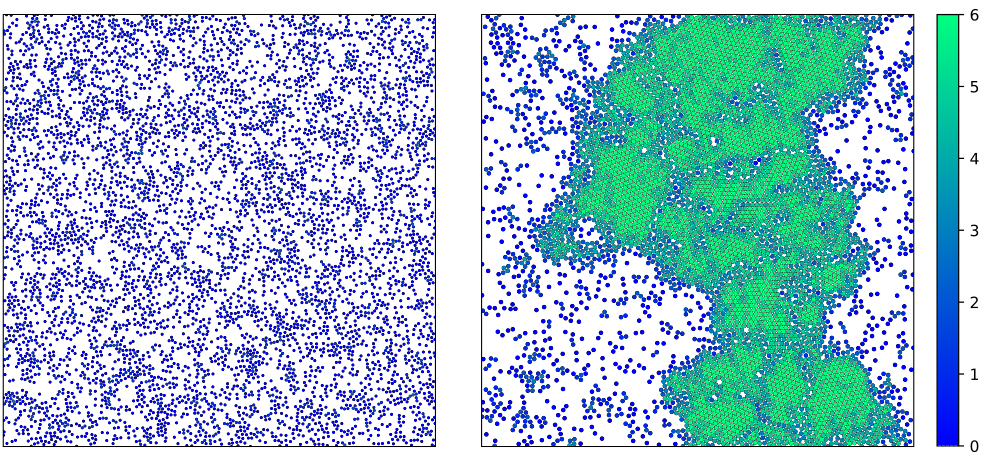}	
\caption{Snapshots of the system containing $N = 6 \times 10^{3}$ active Brownian particles. Each particle is colored according to the number of neighbors within a cut-off distance $C_{r}$. (Left) Dilute regime with packing fraction $\phi = 0.20$ and low Péclet number, $\text{Pe} = 10$. (Right) Dense regime with $\phi = 0.48$ and high Péclet number, $\text{Pe} = 100$, where motility-induced phase separation (MIPS) is visible.
}
\label{fig:1}%
\end{figure}

During this phase transition the system pass from a single phase which recalls an active {\it gas} with an homogeneous particle distribution as shown in Fig.~\ref{fig:1}(left), to a out-of-equilibrium phase, where two phases are distinguishable, one dilute phase which recalls a {\it fluid} and an dense phase or {\it active solid}, as shown in Fig.~\ref{fig:1}(right). The role of $\phi$ and $\text{Pe}$ during the phase transition has been extensively studied in experiments, numerical simulations, and under various theoretical approximations~\cite{stenhammar2013continuum,cates2024active,digregorio2018full,chacon2022intrinsic,cates2015motility}. 

Fixing the rotational diffusion coefficient, $D_r = 1/t_r$, sets the rotational relaxation time of each particle. Then, the self-propulsion speed, and therefore the Péclet number, governs the inter-particle collision rate. This rate grows until a critical threshold $\text{Pe} \simeq 60$ is reached, beyond which particle nucleation or aggregation dominates the system dynamics~\cite{stenhammar2014phase,levis2017active}. Conversely, increasing the packing fraction $\phi$ enhances the likelihood of collisions; steric interactions then reduce the particle velocity and ultimately drive the system into two coexisting phases.

Following Refs.~\cite{norambuena2020understanding,soto2024kinetic,stenhammar2013continuum}, in a dilute active medium the mean free path is
\begin{equation}
\lambda = \left( \frac{\pi}{4} \right)\frac{(1 - \phi)}{\phi},\sigma,
\end{equation}
valid only for $\lambda > C_r$, i.e., when the mean free path exceeds the cutoff interparticle distance. This requirement imposes an upper bound on the packing fraction via the condition $\lambda(\phi_c)=C_r$,
\begin{equation}
\phi < \phi_c=\frac{\pi}{\pi+2^{13/6}}\approx 0.41,
\end{equation}
so that $\phi_c$ provides a good estimate of the crossover packing fraction from the dilute (single-phase) to the dense regime, where the single-phase approximation remains valid.

We perform Brownian dynamics simulations of Eqs.~\eqref{eq1} and \eqref{eq2} with periodic boundary conditions in a square domain of dimensions $L \times L$, using a timestep $\Delta t = 1.5\times10^{-5}t_r$ for numerical integration. Two system sizes are explored: $L_{\text{single}} = 150\sigma$ for the single-phase study and $L_{\text{MIPS}} = 100\sigma$ for the MIPS study. The particle diameter is fixed at $\sigma = 2$, which sets the longitudinal unit scale.
Following Redner \textit{et~al.}~\cite{redner2013structure}, the rotational relaxation time $t_r = 6.6$ defines the natural time unit. Unless otherwise stated, we set the number of particles to $N = 6\times10^{3}$, the rotational and translational diffusion coefficients to $D_r = 0.15$ and $D_t = 0.20$, and the drag coefficients to $\gamma_r = 1$ and $\gamma_t = 0.75\,\gamma_r$. We explore three distinct regimes:  
\begin{enumerate}
    \item \textit{Single-phase ensemble.} Packing fraction $\phi = 0.20$ with Péclet numbers $\text{Pe} \in [10,120]$; total simulation time $T = 30\,t_r$. No MIPS is observed.  
    \item \textit{Phase-separated ensemble.} $\phi = 0.48$, $\text{Pe}=100$, and $T = 60\,t_r$, for which MIPS are observed.  
    \item \textit{Passive reference gas.} Interacting passive Brownian particles at $\phi = 0.20$ with an elevated translational diffusivity $D_t = 100$, a reduced timestep $\Delta t = 1.5\times10^{-6}t_r$, rotational diffusivity $D_r=0$, and $T = 30\,t_r$ with $t_r= 6.6$ as in the previous simulations.
\end{enumerate}

These three regimes unravel the respective roles of activity and density in determining the structural and dynamical properties of the system.

\section{Complex Networks for Active Brownian Particles}
\label{sec.network}
\begin{figure}[ht]
    \centering 
    \includegraphics[width=\columnwidth]{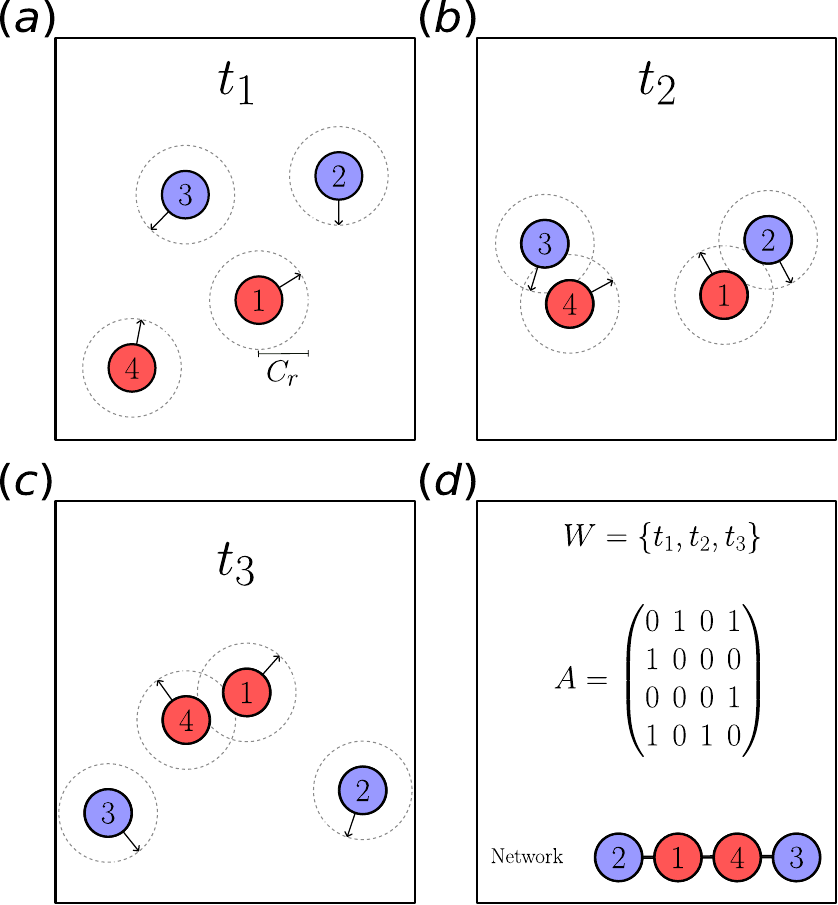}
    \caption{Schematic of particle encounters over a discrete time window. Panels (a--c) show particle positions at three consecutive time steps, with circles of radius $C_r$ indicating each particle’s interaction region. As particles move, reciprocal links form whenever two particles fall within each other’s interaction radius. Panel (d) shows the resulting adjacency matrix $A$ and its graph representation, where $A_{ij}=1$ if particles $i$ and $j$ interacted at any time in $W=\{t_1,t_2,t_3\}$, and $A_{ij}=0$ otherwise.}
    \label{fig:2}%
\end{figure}

A network is defined as a group of nodes connected by edges, where the connections can be either directed or undirected. In our model, each node corresponds to an individual ABP, and their interactions define the connections within a time window $W$: two nodes are connected if $r_{ij} \leq C_r$ within a given time window. A time window is defined as several consecutive time steps, expressed in terms of $t_r$, as schematically shown in Fig.~\ref{fig:2}. Based on this criterion, we constructed a time-dependent network by tracking the dynamics of each particle across the regimes previously described, using time windows of increasing widths ranging from very short intervals ($W = 0.75t_r$) to much longer ones ($W = 30t_r$). However, the phase-separated (MIPS) case requires a slightly different approach to analyze each phase separately. 

In the MIPS regime, we use the DBSCAN algorithm~\cite{DBSCAN1996}, which relies on two parameters: $\epsilon_s$, the maximum distance between two particles for one to be considered within the neighborhood of the other, and $m_s$, the minimum number of neighboring particles required to define a cluster. We set $\epsilon_s = C_r$ and $m_s = 4$ throughout the simulations. Using the DBSCAN method, it is possible to identify each particle as belonging either to the large cluster (dense phase) or not (dilute phase) at each time step. This classification allowed us to analyze the properties of each phase separately.

\subsection{Adjacency matrix}
\label{sec:admatrix}

The adjacency matrix $A$ is the building block for representing a complex network since it encapsulates information about node connections. Temporal connections between ABPs are considered if $r_{ij} \leq C_r$, then $A_{ij} = A_{ji} = 1$; otherwise, $A_{ij} = 0$. Thus, we obtain an undirected network in a given time window. Specifically, if $r_{ij} \leq C_r$, then $A_{ij} = A_{ji} = 1$; otherwise, $A_{ij} = 0$, thus we obtain an undirected network. 

In the MIPS regime, a distinct procedure is used to compute $A_{ij}$. When analyzing each phase separately, we impose two additional conditions: since a particle may transition between phases within a single time window (though not at the same time step), particles $i$ and $j$ are considered connected only if they belong to the same phase at a given time step. In addition, each particle is assigned to the adjacency matrix of the dense (dilute) phase if it spends more than half of the corresponding time window in that phase. We compute key network metrics from these matrices, including the degree distribution, global clustering coefficient, and average path length. These quantities provide valuable insight into the structure and organization of the underlying interaction network.

\subsection{Degree distribution $P(k)$}
\label{sec:degreedistribution}
To characterize the structure of a complex network, we begin by analyzing the degree distribution. The degree $k_i$ of a node $i$ is defined as the number of unique encounters the corresponding particle experiences within a given time window, and is defined as
\begin{equation}
    k_i = \sum_{j=1}^{N} A_{ij} = \sum_{j=1}^{N} A_{ji}
    \label{eq5},
\end{equation}
where the adjacency matrix elements $A_{ij}$ are computed as explained in Sec.~\ref{sec:admatrix}. The degree distribution $P(k)$ read as
\begin{equation}
    P(k) = \frac{n_k}{N},
    \label{eq6}
\end{equation}
with $n_k$ representing the total number of particles with degree $k$, and $N$ the total number of particles in the system. The above metric can capture the statistical properties useful to classify the resulting networks, for instance, as random (Erdős-Rényi~\cite{erdos1960}), scale-free (Barabási-Albert~\cite{barabasi1999}), or of another type. More importantly, changes in $P(k)$ can signal structural transitions in the network~\cite{pasten2018}, which will be useful to characterize different ABP regimes explored in this work. From an active matter perspective, $P(k)$ encodes how particles interact over time and reflects the evolving nature of connectivity in the system regarding the P\'eclet number and packing fraction. Depending on the topology of a network, the degree distribution could show different behaviors~\cite{Erdos,barabasi1999,albert2002statistical,newman2003}. 

\subsection{Network Global clustering coefficient}
\label{sec:clustering}
Another key property of complex networks is the small-world phenomenon~\cite{watts1998}. Watts and Strogatz introduced a network model that exhibits small-world characteristics, namely a low average path length combined with a high clustering coefficient. These features distinguish small-world networks from random graphs and regular lattices, which lack a balance of local and global connectivity. The local clustering coefficient (defined for each node) and the global clustering coefficient (defined for the entire network) can be analyzed to quantify this behavior. The clustering coefficient~\cite{Newman_2000,newman2003} measures the tendency of nodes to form tightly connected groups, with higher values indicating a greater propensity for clustering. The global coefficient, in particular, captures the overall tendency of the network to form such clusters, and is defined as:

\begin{equation} 
C = \frac{\displaystyle{\sum_{i,j,k} A_{ij} A_{jk} A_{ki}}}{\frac{1}{2} \displaystyle{\sum_i k_i (k_i - 1)}},
 \quad \quad 0 \leq C \leq 1.
    \label{clustering}
\end{equation}

Note that the clustering coefficient is a quantity accumulated over time windows; therefore, in this context, a clustering coefficient close to 0 represents a complex network without triplets, and a clustering coefficient similar to 1 represents a fully connected complex network.

\subsection{Average path length}
To further characterize small-world behavior, we evaluate the average path length, defined as the mean shortest distance between all pairs of nodes in the network~\cite{Newman_2000}. In small-world networks, both high clustering and low average path length coexist, distinguishing them from purely random or regular structures. The average path length is given by:

 \begin{figure*}[ht!]
    \centering 
    \includegraphics[width= 1 \textwidth]{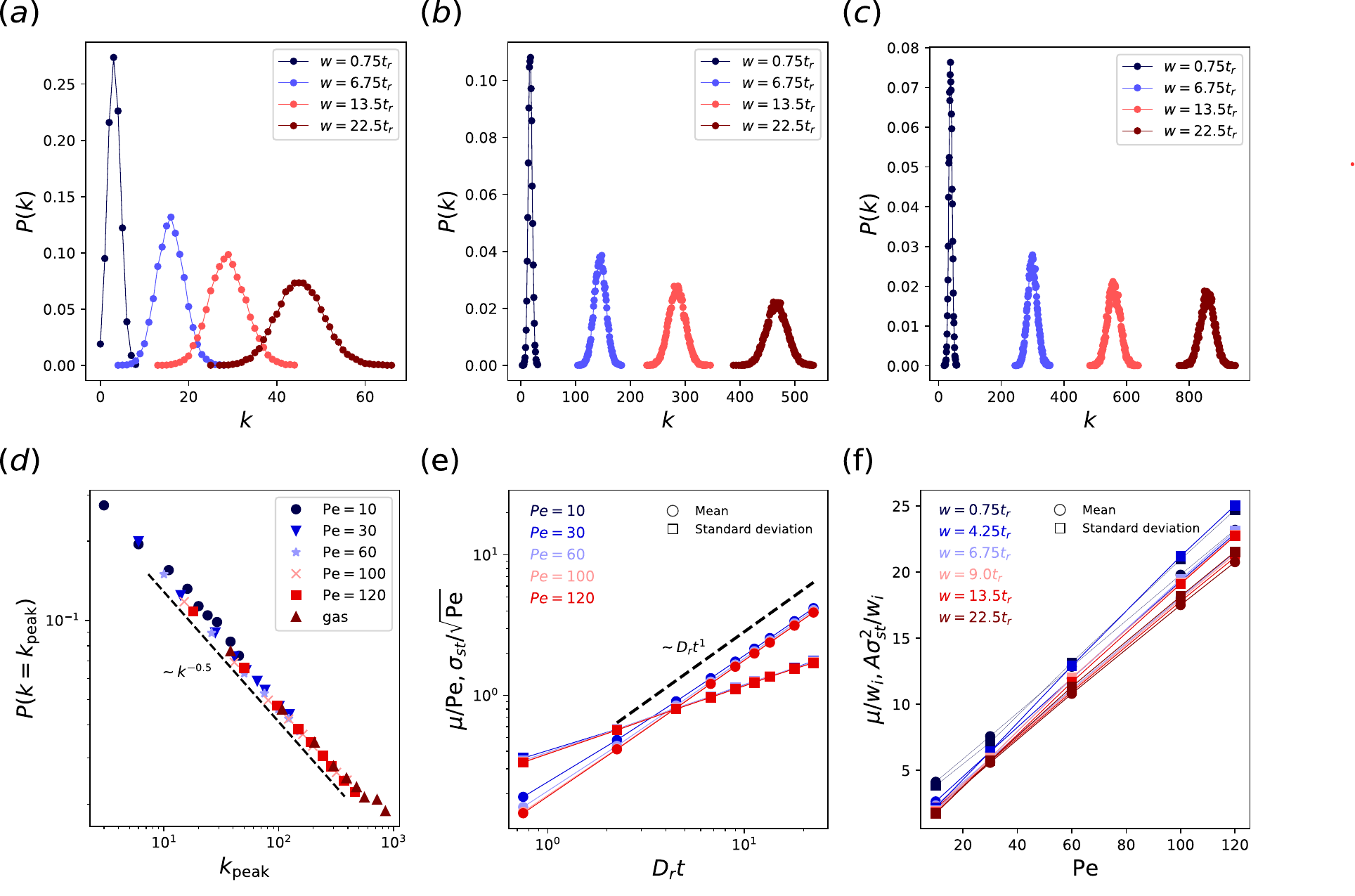}
\caption{Top: Degree distributions of systems in the single-phase regime for $\phi=0.2$ and different Péclet numbers: $ \text{Pe} = 10 $ and $ \text{Pe} = 120 $ for panels (a) and (b), respectively. The degree distributions are truncated when $P(k)=0$. Color indicates the width of the time window considered, transitioning from darker blue (shorter window) to darker red (longer window) as the window increases. Panel (c) shows the degree distribution of a gas of passive Brownian particles with a  higher diffusion coefficient. Similarly, color indicates time window width, ranging from dark blue (short) to dark red (long). Bottom: 
Panel (d) shows the maximum value of $P(k_{\text{peak}}) \sim k_{\text{peak}}^{-\gamma}$ that decays as a power law with a critical exponent $\gamma = 0.5$, for different Péclet number $\text{Pe} \in [10, 120] $ at different time windows, including the passive Brownian gas. Panel (e) and (f) shows the  relation followed by $\mu(\mathrm{Pe},W)=0.18\;\mathrm{Pe}\;W_i$ and $\sigma_{\mathrm{st}}(\mathrm{Pe},W)=\sqrt{0.18\;\mathrm{Pe}\;W_i/A}$ with $A=2\pi \Lambda^2$, in terms of the time windows $W_i=D_rt$ and $\mathrm{Pe}$ respectively.}
\label{fig:3}%
\end{figure*}

\begin{equation}
l_G = \frac{1}{N(N - 1)} \sum_{i \neq j} d(v_i, v_j),
\label{lg}
\end{equation}

where $d(v_i, v_j)$ denotes the shortest number of steps between nodes $v_i$ and $v_j$. In the Watts–Strogatz model~\cite{newman_watts1999}, this length scale depends on the rewiring probability, illustrating how local changes in connectivity can produce global shortcuts. Together with the clustering coefficient, the average path length is a key indicator of small-world properties in active-matter networks. When $l_G$ remains small compared to the network size, it implies that any two particles can be connected through only a few intermediaries, reflecting efficient information or interaction flow in the system.

\begin{figure}[ht!]
    \centering 
    \includegraphics[width= 1.1\columnwidth]{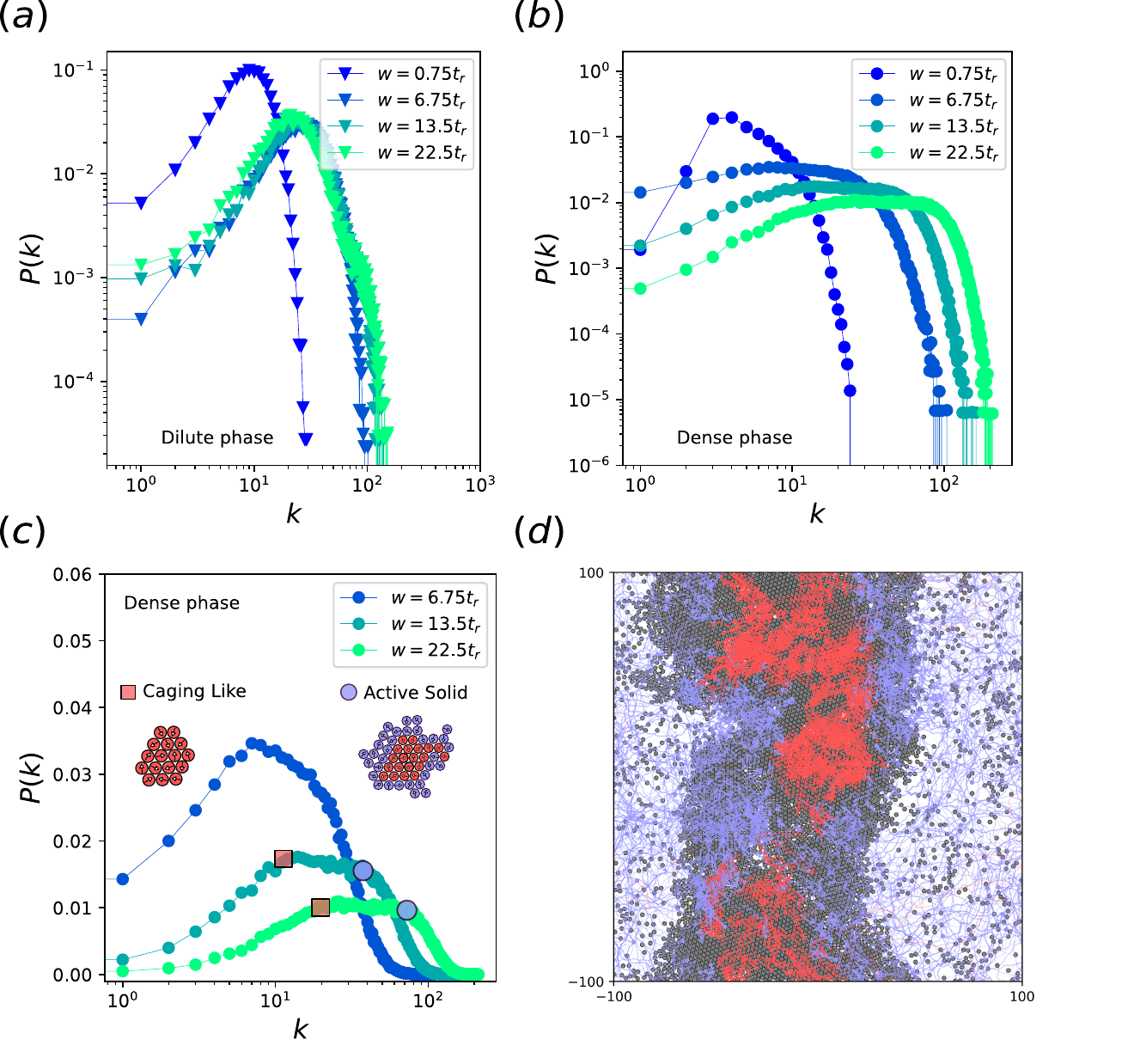}
    \caption{Log-log plots of the time‐averaged degree distribution $P(k)$ for four observation windows $W_i$ during motility-induced phase separation at $\phi = 0.48$ and $\text{Pe}=100$. (a) Dilute phase:  $P(k)$ is non-Gaussian, with the highest probability at $k\sim20$ and an abrupt drop once $k\!\approx\!100$, the value corresponding to a particle that connects to a full row of neighbours across the box. The qualitative shape is unchanged by the choice of $W_i$, but the amplitudes differ slightly. (b) Dense phase: At early times (blue), the distribution rises sharply at $k=6$, then flattens into a plateau for $6 \lesssim k \lesssim 100$. The plateau becomes more pronounced for larger windows (green). (c) Semi-log average degree distribution during the dense phase, for three time windows. The plateau has two equally probably limits: one near $k \approx 10$, associated with particle caging in quasi-hexagonal structures (red square), and another near $k \approx 100$, linked to long-range connections within active solid regions (red circle). (d) Particle trajectories in the dense phase for a time window $W = 13.5t_r$. Particles with a small number of connections (red square in (c)) are shown in red, while those with a large number of connections (purple circle in (c)) are shown in purple. Gray particles represent a static image of the formed cluster.}
\label{fig:4}%
\end{figure}

\section{Results and Discussion}
We now build the complex network described in Section~\ref{sec.network}. The adjacency matrix is used to construct the degree distribution in two cases: (A) a single phase with fixed $\phi =  0.2$ and several $\text{Pe}=[10,30,60,100,120]$, and (B) a two-phase system with $\phi = 0.48$, and $\text{Pe}=100$.

\subsection{One-phase region: Random Graph}

In the single-phase region, the node degree distribution follows a Gaussian function for every time window $W_i$ (see Fig.\ref{fig:3}(a) and (b), where we truncate the tails with zero probability to make the panels clearer). To benchmark these results, we simulate a \textit{pure gas} in which particles undergo only translational diffusion with $D_t = 100$, i.e., without self-propulsion or rotational diffusion. Figure~\ref{fig:3}(c) shows that its degree distribution coincides with that of the active gas during the single phase. We capture this \textit{active gas} behavior with the phenomenological degree distribution,
\begin{eqnarray} \label{GuassianDistribution}
 P(k, \text{Pe}, W) = {\frac{1}{ \sigma_{\rm st}  \sqrt{2\pi}}} \mbox{exp}\left[ - \frac{1}{2}\left( \frac{k-\mu}{\sigma_{\rm st}}\right)^2 \right],
 \end{eqnarray}
where the mean degree $\mu(\phi,\mathrm{Pe},W)$ and the standard deviation $\sigma_{\mathrm{st}}(\phi,\mathrm{Pe},W)$ depend (in principle) on the packing fraction $\phi$, the Péclet number $\mathrm{Pe}$, and the observation window $W$. We identify the most probable number of connections a particle makes in a time window $k_{\text{peak}}$, by identifying the peak of the Gaussian distribution $P(k = k_{\text{peak}})$. We plot this quantity in Fig.~\ref{fig:3}(d), where we, surprisingly, observe a collapse of the curves for all $\mathrm{Pe}$,

\begin{equation}
P(k_{\text{peak}})=\Lambda/k_\text{peak}^{0.5}
\label{pmax}
\end{equation}

with $\Lambda=0.47$, by fitting Eq.\eqref{GuassianDistribution} to the degree distributions obtained for simulations in the single-phase regime, and extracting $\mu(\mathrm{Pe},W)$ and $\sigma_{\mathrm{st}}(\mathrm{Pe},W)$, we observe that $P(k)$ is maximum when $\mu = k_{\text{peak}}$. Evaluating Eq.\eqref{GuassianDistribution} at this point yields, $P(k_{\text{peak}})=1/\sigma_{\mathrm{st}}\sqrt{2\pi}=\Lambda/\sqrt{\mu}$, from this, we obtain the relation $\mu \propto \sigma_{\mathrm{st}}^2$. This scaling reveals a fundamental statistical property: the mean degree of the distribution is proportional to its variance. Physically, this implies that as nodes increase their average number of connections, the fluctuations in their connectivity grow even faster. Moreover, this relation recovers Eq.~\eqref{pmax}.

In terms of the time dependence of $\mu(\mathrm{Pe},W)$ and $\sigma_{\mathrm{st}}(\mathrm{Pe},W)$, we also observe a collapse of these quantities when compared across $\mathrm{Pe}$, as shown in Fig.\ref{fig:3}(e). This reflects the relations $\mu(\mathrm{Pe},W) \propto \mathrm{Pe}\;W$ and $\sigma_{\mathrm{st}}(\mathrm{Pe},W) \propto \sqrt{\mathrm{Pe}\;W}$. Finally, Fig.\ref{fig:3}(f) presents these relations explicitly in terms of $\mathrm{Pe}$, where we find that short time windows deviate from the proportional scaling, whereas at larger time windows the curves follow the same proportionality.

\subsection{Two-phase region}

Within the MIPS region characterized in Ref.~\cite{redner2013structure}, we reproduce phase separation at packing fraction $\phi = 0.48$ and Péclet number $\mathrm{Pe} = 100$. Following the adjacency matrix construction outlined in Sec.~\ref{sec:admatrix}, we analyze the resulting dilute and dense phases (see Supp. Movie 1~\cite{movies}). Once both phases are fully established, we generate the complex network approximately after $t_{\text{run}} \sim \text{30} t_r$.

In Figure~\ref{fig:4}, we present the log-log average degree probability distribution for different time windows $W$, with blue dots representing the shorter time windows and green dots representing the longer ones. We observe that the distribution for the dilute phase (Fig.\ref{fig:4}(a)) does not change qualitatively with increasing window length; only slight quantitative differences appear at higher degrees, with the maximum of the distribution at $k \sim 20$ connections. Although this dilute phase might be expected to exhibit gas-like behavior, we find that after the phase transition (during MIPS) the degree distribution is no longer Gaussian, in agreement with the qualitative change reported by Nayak et al.\cite{nayak2025diffusion} in their study of particle displacements during MIPS.

We now focus on the degree distribution $P(k)$ for the dense phase (Fig.~\ref{fig:4}(b)). We observe that low-degree values ($k<6$) become less likely as the time window increases. For shorter time windows, such as $ = 6.75\,t_r$ (in dark blue dots), the distribution exhibits a peak around the formation of the hexatic cluster $k=6$. As the time window increases, this peak decreases and the degree distribution flattens to a plateau, an indication of clustering behavior, where particle encounters with degrees in the range $k \approx 10$ to $\approx 100$ become equally probable. Beyond this range ($k>100$), the distribution decays, reflecting rare events in which a particle within the cluster travels across its boundary, forming a large number of transient connections. This behavior is similar to what is observed in the dilute phase.

To understand the plateau reached during the dense phase we display the semi-log average degree distribution for three time windows (see Fig.~\ref{fig:4}(c)): an early one at $W = 6.75t_r$ (in blue) and a later one at $W = 22.5t_r$ (in green). 

The plateau has two marked limits, one at $k \approx 10$, corresponding to hexagonal packing, and another around $k \approx 100$, associated with the connectivity of the system size scale. Although these values are close to expected structural configurations, their slight shifts suggest complex dynamics. At small degrees, the first limit is attributed to events of particles performing a \textit{caging-like} process in which particles are trapped in a hexagonal structure and occasionally escape, thus increasing their number of contacts (see the red square).
The second limit at higher degrees corresponds to a less probable scenario at shorter times (blue dotted curve), where the particles undergo \textit{active solid} like behavior~\cite{digregorio2018full,goswami2025yielding, caporusso2020motility}. In this regime, particles maintain extended contact as they move along the borders of grain boundaries or travels along larger channels due to gas bubble ruptures within the cluster~\cite{caporusso2020motility, umemura2025density}, resulting in a larger number of connections (see the red circle).
To visualize these two behaviors, in Fig.~\ref{fig:4}(d) we show, as a guide to the eye, a static image of the cluster (gray dots), with the trajectories of ten particles performing a caging-like behavior shown in red, and the trajectories of eighty particles exhibiting an active-solid behavior shown in purple (see Supp. Movie 2~\cite{movies}). Both sets of particles diffuse over the same time interval. We have demonstrated that, by analyzing the degree distribution, it is possible to identify local microphases within the dense phase, in a manner similar to the approach used by Paoluzzi et al.\cite{paoluzzi2022motility} for a polydisperse collection of ABPs.

\subsubsection{Role of the global clustering coefficient}

\begin{figure}[ht!]
    \centering 
    \includegraphics[width= 1 \columnwidth]{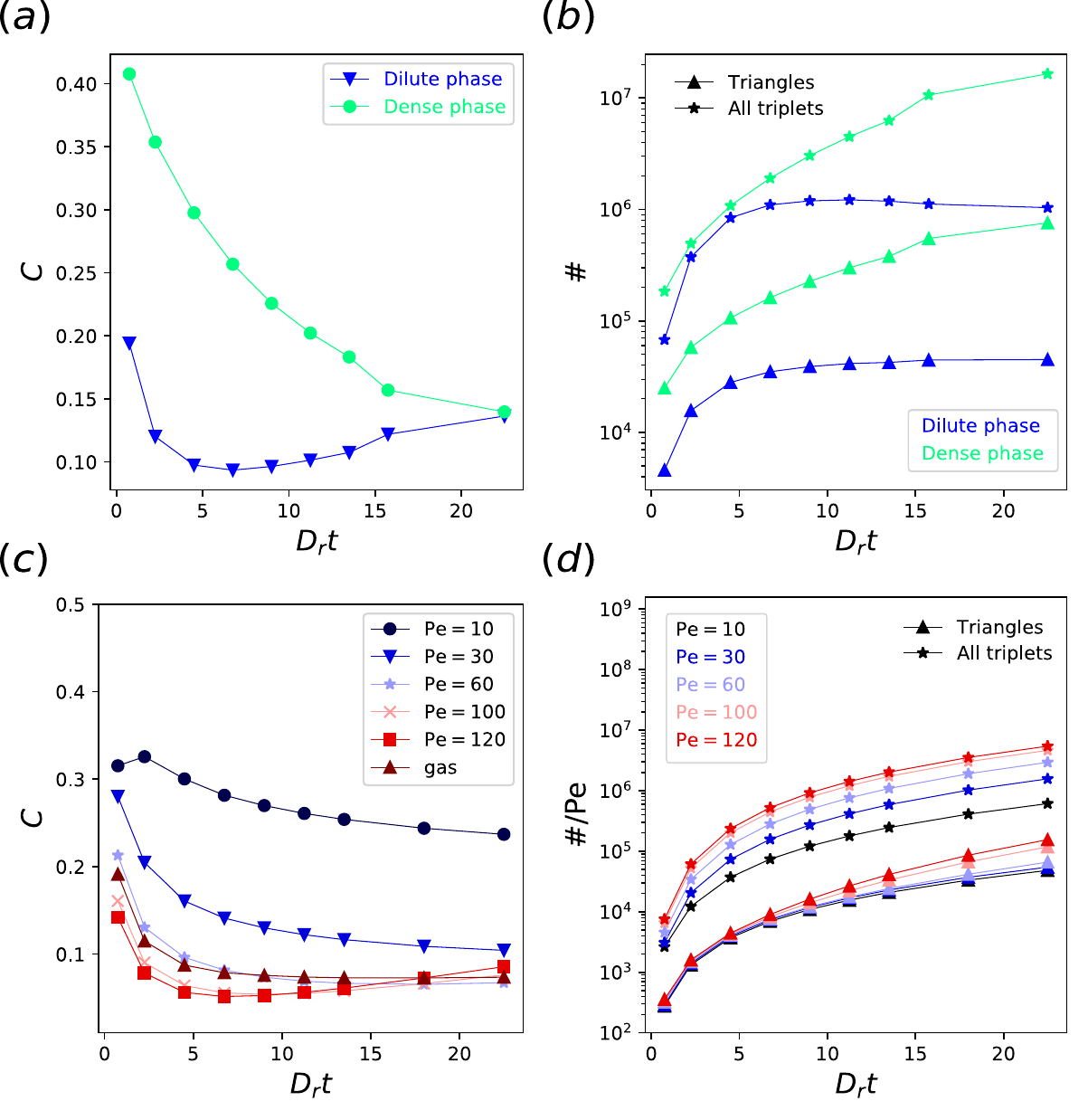}
    \caption{Global clustering coefficient ($C$) and total number of triplets and triangles $\#$ for the two-phase and single-phase regions across different time windows, respectively.
    (a)-(b) System during MIPS: Blue triangles represent the dilute region, green dots correspond to the dense region. 
    (c)-(d) One phase case: The global clustering coefficient $C$ and the number of triangles and triplets $\#/$ for different P\'eclet numbers $\text{Pe}$.} 
\label{fig:5}%
\end{figure}

The global clustering coefficient ($C$), along with the total number of triplets and triangles, is shown in Fig.\ref{fig:5} for both the two- and single-phase regions across different time windows. A \textit{triplet} is defined as a group of three nodes connected by either two (open triplet) or three (closed triplet) undirected edges. A \textit{triangle} corresponds to three mutually connected nodes, i.e., a fully connected triplet, which includes three closed triplets—one centered on each node. Under these definitions, the global clustering coefficient is given by the ratio between the total number of closed triplets (three times the number of triangles) and the total number of triplets (see Section\ref{sec:clustering}).

In Figs.~\ref{fig:5}(a)--(b), we present results for the two-phase system, where blue triangles represent the dilute region and green dots correspond to the dense region. The dilute phase exhibits a rapid decay in $C$ and shows a non-monotonic trend. The dense phase starts with $C \sim 0.4$, corresponding to a hexagonal configuration, and slowly decays toward a steady state. The total number of triplets (indicated by stars) and the total number of triangles (indicated by triangles) grow rapidly in both the dilute and dense phases. While the number of triplets and triangles in the dilute phase eventually reaches a steady state, in the dense phase it continues to grow, highlighting a key feature of active systems: the prevalence of three-particle encounters in the dense phase.

Fig.\ref{fig:5}(a) shows a collapse to the value $C \sim 0.15$ for both phases at large time windows. Unlike the $C \sim 0.4$ limit, which has a clear geometrical origin in hexagonal packing, the $C \sim 0.15$ value does not correspond to an ordered structure. Instead, it can be interpreted as the average geometric probability of forming closed triplets in a disordered contact network of hard particles, where the exclusion volume and local density correlations increase the likelihood of incidental triangular configurations. This value appears to be governed by the Péclet number, since the local packing fractions in the dilute and dense phases are not the same. Importantly, the system becomes diffusive at longer times\cite{nayak2025diffusion,umemura2025density}, as evidenced by the continued growth in the number of triangles and triplets, with triangles being an order of magnitude more abundant in the dense phase.

To further investigate the role of the Péclet number in the clustering coefficient, in Fig.\ref{fig:5}(c) we present $C$ for the single-phase regime. We observe that $C$ decreases for more persistent particles, i.e., at larger Péclet numbers. This occurs because, when $\text{Pe} \geq 60$, the system undergoes a behavioral change: increased particle persistence enhances both the likelihood of three-particle encounters and the duration of their interactions, due to the appearance of large density fluctuations in the system\cite{caporusso2020motility,redner2013structure}.

Conversely, less active particles at small Péclet numbers diffuse more slowly, as described by the relation $D_{\text{active}} \approx u_0^2 t_r / 2$\cite{stenhammar2014phase,soto2025self}. As a result, particles at high Péclet numbers are more likely to form clusters over time, producing a non-monotonic trend similar to that observed in the dilute phase in Fig.\ref{fig:5}(a). This trend is also reflected in the number of triangles and triplets in Fig.~\ref{fig:5}(d), with the number of triangles increasing at a higher rate for larger Péclet numbers at longer time windows.

Although the number of triplets remains higher than the number of triangles, triangle formation becomes more probable as particle persistence increases. Figure~\ref{fig:5}(d) also shows the scaling of the number of triangles with the Péclet number, demonstrating that particle persistence largely determines the clustering coefficient $C$.

\subsubsection{Role of the average path length}
\begin{figure}[ht!]
    \centering 
    \includegraphics[width= 1\columnwidth]{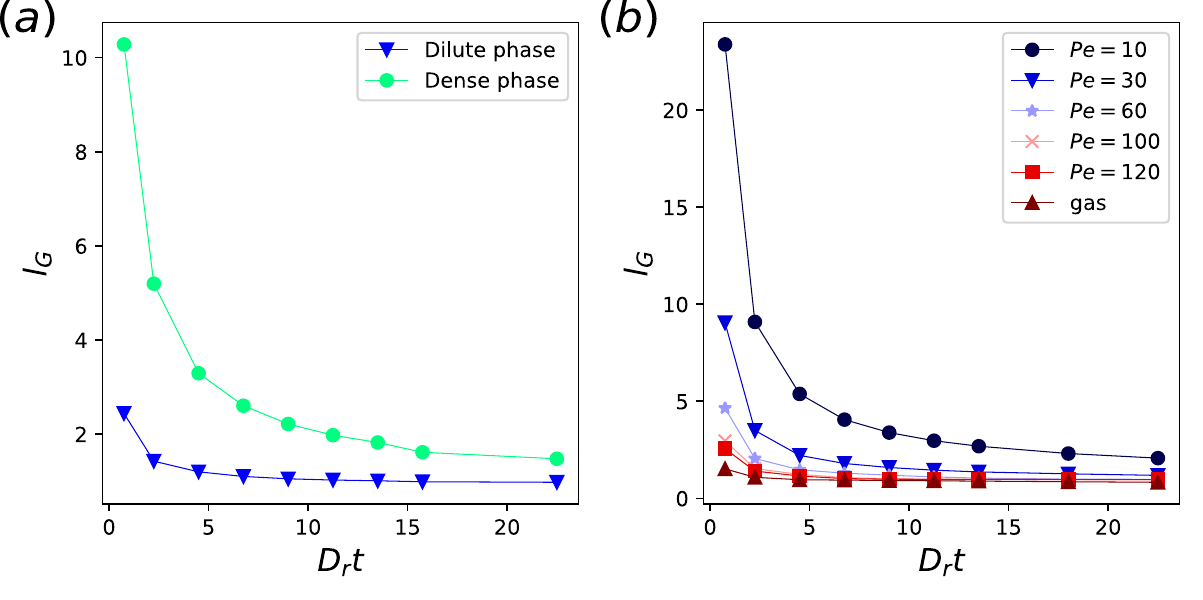}
    \caption{The average path length $l_G$ as the time window increases. Left: for the two-phase system, and Right: for the single phase at different Péclet numbers.} 
\label{fig:6}%
\end{figure}

The average path length for the dilute phase during MIPS and single-phase systems with higher diffusion coefficients, referred to as the {\it gas} case, and larger Péclet numbers (shown in red and light blue in Fig.~\ref{fig:6}) remains nearly constant and exhibits lower values of $l_G$ as the time window increases. Moreover, these systems display a collapse for $D_r t_r > 15$. In contrast, systems with less active particles ($\text{Pe} = 10, 30$) exhibit a significantly larger average path length that gradually decreases with increasing time window. A similar trend is observed in the dense phase during MIPS. These results suggest that a long average path length is a hallmark of systems composed of particles with low mobility, whereas systems with higher diffusivity or those in a {\it gas}-like state exhibit shorter average path lengths, reflecting more transient and spatially homogeneous interactions.

\section{Conclusions}

In this work, we studied the behavior of Active Brownian Particles (ABPs) by constructing and analyzing complex networks based on particle contacts over finite-time windows. We examined how network structure evolves with time window length in both the single-phase regime and the Motility-Induced Phase Separation (MIPS) region. Tracking key network metrics—degree distribution, clustering coefficient, and average path length—revealed how time-accumulated interactions generate distinct network topologies in each regime. This approach provides a universal framework for characterizing interaction structures and enables direct comparisons across systems.

In the single-phase region, Gaussian degree distributions emerge, consistent with random contact networks. By combining simulations with phenomenological analysis, we uncovered a robust scaling law: the mean degree is proportional to the variance, $\mu \propto \sigma_{\mathrm{st}}^2$, with both quantities increasing proportionally with time window $W$ and Péclet number $\mathrm{Pe}$. This scaling explains the universal collapse of the maximum value of $P(k)$, independent of $\mathrm{Pe}$. The single-phase regime is further characterized by low clustering coefficients and small average path lengths, reflecting random, short-lived contacts.

In the MIPS regime, ABPs spontaneously segregate into dense and dilute regions due to persistent motion and excluded-volume interactions. Constructing networks separately for each phase revealed qualitative differences: the dilute phase retains a stable degree distribution with a marked peak at $k \gtrsim 10$, while the dense phase evolves from hexagonal caging ($k \sim 6$) at short windows to broader distributions at longer windows, with connectivity reaching $k \approx 100$. This growth reflects the active-solid nature of the dense phase, where microphase dislocations, bubbles, and dynamic reconfigurations~\cite{digregorio2018full,goswami2025yielding,caporusso2020motility} enable particles to travel long distances within clusters.

Our results demonstrate that complex network analysis not only captures dynamic phase transitions in out-of-equilibrium systems but also reveals hidden scaling relations and microstructural signatures within each phase. This framework is transferable to related systems—such as granular media, animal groups, or human crowds—by characterizing phases through network metrics. Beyond characterization, it offers tools to monitor and potentially control clustering in both living and synthetic active matter, with implications for processes such as disease transmission. Altogether, these findings establish complex network analysis as a powerful and generalizable methodology for probing phase behavior in active matter and other non-equilibrium systems.
\section*{Acknowledgments}
The authors are grateful for the helpful feedback given by Pablo de Castro and Aparna Baskaran. I.S and F.G.-L. have received support from the ANID – Millennium Science Initiative Program – NCN19 170, Chile. F.G.-L. was supported by Fondecyt Iniciación No.\ 11220683. A.N. acknowledges the financial support from the project Fondecyt Iniciaci\'on No. 11220266.


\begin{thebibliography}{61}
\bibitem [{mov()}]{movies}%
  \BibitemOpen
  \href@noop {} {\bibinfo  {journal} {See Suplemental Material at [] for Movie
  1, Two-phase system: Blue indicates particles that, within this time window,
  belong to the dense phase, and green indicates particles that belong to the
  dilute phase, and Movie 2, Particles in the dense phase performing
  caging-like behavior in red, and those exhibiting active-solid behavior in
  purple, within a given time window}\ }\BibitemShut {NoStop}%
\end{thebibliography}
%

\end{document}